\documentstyle[aps,pre,epsf]{revtex}

\begin{document}
\draft
\title{{\bf Do correlations create an energy gap in electronic bilayers?\\ Critical analysis of different approaches}}
\author{ J. Ortner}
\address{
{\it Institut f\"ur Physik, Humboldt Universit\"{a}t zu Berlin, 
Invalidenstr. 110, D-10115 Berlin, Germany}}

\date{\today}
\maketitle

\begin{abstract}
This paper investigates the effect of correlations in electronic bilayers on the longitudinal collective mode structure. We employ the dielectric permeability constructed by means of the classical theory of moments. It is shown that the neglection of damping processes overestimates the role of correlations. We conclude that the correct account of damping processes leads to an absence of an energy gap.   
\end{abstract}

\pacs{73.20.Mf,71.45.Gm,73.50.Mx,52.35.Fp}


Layered electronic systems have been of intense interest recently. Such systems can now be routinely fabricated in semiconductors with well-controlled system parameters. We focus in this paper on the analysis of the collective modes of electronic bilayers. These systems consist of two quasi-two-dimensional (2D) layers of electron liquids with electron density $n_s$, separated by a distance $d$ comparable to the interparticle distance $a=(n_s/\pi)^{1/2}$ within the layers. 
Quasi-two-dimensional here means that the electrons have quantized energy levels along one dimension, but are free to move in two dimensions. A well known example of 2D electron system are electrons confined in the vicinity of a junction between a semiconductor and isolators [in a MOSFET structure] or between layers of different semiconductors [in heterojunctions]. In these systems electrons are confined near the surface by an electrostatic field; in electronic bilayer systems two electron layers are separated in a double quantum well. 
Another class of layered systems, electronic superlattices, consisting of a large number of identical electronic layers, show a similar behavior and are beyond the scope of this paper. High $r_s$ ($r_s =a/a_B$; $a_B=\varepsilon_s \hbar/e^2m^*$ being the effective Bohr radius) values are now available in 2D layers \cite{Shapira96} and the technique should be available to fabricate relativley high $r_s$ bilayers also. One expects that at high enough $r_s$ values ($r_s > 20$ \cite{Swierkowski91}) a bilayer crystallizes into a Wigner lattice. Here, we restrict our considerations to the strongly coupled liquid phase bilayer systems. There are some theoretical investigations in this strongly-coupled liquid phase regime, both on the static and the dynamic level \cite{ZT88,Gold92,SNS93,KRG93,SSN94,ZM94,VKB96,VKB97}. 

One of the issues under consideration is the analysis of collective excitations in bilayers. First we recall the Random phase approximation (RPA) result \cite{Fetter74} applicable in the weak coupling regime $r_s \ll 1$. In the RPA there are two longitudinal modes: an in-phase and an out-of-phase mode. In the first mode the two layers oscillates in phase and the resulting dispersion relation is similar to that of an isolated 2D layer; for $k \to 0$ the eigenmode behaves as $\omega \sim \sqrt{k}$. In the out-of-phase mode the oscillation phase of the two layers differs by $\pi$ and we deal with an acoustic mode $\omega \sim k$ as $k \to 0$. However, in the strong coupling regime $r_s \gg 1$ the collective modes structure is strongly affected by particle correlations. The effects of correlation beyond RPA can be recasted into the local field correction. Early approaches focused on intralayer correlations but ignored or suppressed interlayer correlations \cite{SSN94}. 
There have been two basic lines which also take into account the interlayer correlations. One line uses a low frequency (static) approximation for the local field correction. This can be done by applying the Singwi-Tosi-Land-Sjolander (STLS) approximation \cite{STLS68} to the bilayer problem \cite{ZT88,Gold92,SNS93}. The other line uses a high-frequency local field correction. The analytical most simple expressions can be obtained using the quasilocalized charge (QLC) approximation \cite{KG90}. Related studies to the bilayer system have been done in Refs.\cite{KRG93,VKB96}. 
Examiming the effect of the particle correlations on the collective excitations structure the two methods have arrived at different results. In particular, the QLC method predicts the occurence of a finite energy gap ($\omega > 0$ for $k = 0$); however, no such energy gap appears in the calculations based on the STLS approximation. In Ref. \cite{KG98} the formal reasons leading to the different results are clarified. It has been shown, that in the STLS formalism the local field correction vanishes as $k \to 0$ and as a result we arrive (qualitatively) at the RPA dispersion expression without an energy gap. On the other hand, the high-frequency QLC local field correction does not vanish at $k = 0$ and this leads to the occurence of the energy gap. This paper is adressed to the question whether interlayer correlations create an energy gap in bilayer systems. 

The bilayer, consisting of two 2D electron layers of area $A$ embedded in a neutralizing background, and separated by distance $d$ can be mapped onto a two-component two-dimensional system \cite{KRG93}. In the following we neglect the layer thickness, the influence of metallic electrodes and the difference between the dielectric constants of different semiconductors. Then the corresponding interaction potentials are
\begin{eqnarray} \label{potential}
\varphi_{11}(k)&=&\varphi_{22}(k)=\frac{2\pi e^2}{k} \,\,,\nonumber\\
\varphi_{12}(k)&=&\frac{2\pi e^2}{k}e^{-kd} \,\,.
\end{eqnarray}
The two component system may be described by a matrix formalism in species space \cite{KG84,ORT94}. However, in the present case of two identical electron layers the corresponding matrices diagonalizes and it is much easier to investigate the scalar dielectric functions for the in-phase and out-of-phase motions, $\varepsilon_{in}$ and  $\varepsilon_{out}$ respectively,
\begin{equation} \label{epsilon}
\varepsilon^{-1}_{\alpha}({\bf {k}},\omega)=1+\varphi_{\alpha} \chi_{\alpha}({\bf {k}},\omega)\,\,\,\,, ~ \alpha={\rm in,out} \,\,.
\end{equation}
where $\varphi_{{\rm in,out}}=2\pi e^2 (1\pm e^{-kd})/k$ are the interaction potentials for the in- and out-of-phase motions, respectively, and
\begin{equation} \label{chi}
\chi_{\alpha}({\bf {k}},\omega)=(\hbar A)^{-1} \int \, d{\bf r} \,  \int_{-\infty}^{0}\,dt\, <[n_{\alpha}({\bf r},t),n_{\alpha}(0,0)]> e^{i{\bf k}\cdot {\bf r}+i\omega t}\,  \,\,,
\end{equation}
are the density response functions. Here $n_{{\rm in,out}}({\bf r},t)=n_1({\bf r},t) \pm n_2({\bf r},t)$, $n_i({\bf r},t)$ ($i=1,\,2$) being the electron number density operator of the $i$th layer, $[A,B]$ is the commutator of operators $A$ and $B$, and $<A>$ is the equilibrium average of operator $A$.

The calculations based on the QLC formalism \cite{KG90,KRG93} lead to the following expression for the bilayer dielectric permeabilities,
\begin{eqnarray} \label{QLCeps}
\varepsilon^{-1}_{{\rm in}}({\bf {k}},\omega)&=&1 + \frac{w^2_{1}(k) (1+e^{-kd})}{\omega^2 -w^2_{1}(k) (1+e^{-kd}) - (D_{11}+D_{12})} \,\,, \nonumber\\ 
\varepsilon^{-1}_{{\rm out}}({\bf {k}},\omega)&=&1 + \frac{w^2_{1}(k) (1-e^{-kd})}{\omega^2 -w^2_{1}(k) (1-e^{-kd}) - (D_{11}-D_{12})} \,\,, 
\end{eqnarray}
where $w_1(k)=(2 \pi e^2 n_s k/m)^{1/2}$ is the 2D plasma frequency of an isolated electron layer with surface number density $n_s$. The positive magnitudes $D_{11}$ and $D_{12}$ take into account the intra- and interlayer Coulomb correlations between the electrons, and are expressable via the Fourier transform of the pair correlation functions $h_{11}(k)$ and $h_{12}(k)$, so that \cite{KRG93,VKB96,KG98}
\begin{eqnarray} \label{D11}
D_{11}(k)&=& - \,\, w^2_1(k) \sum_{{\bf q}} \frac{\left({\bf k} \cdot {\bf q} \right)^2}{k^3 q} \,\, \left\{ h_{11}(|{\bf k} - {\bf q}|) - h_{11}(q) - e^{-qd}h_{12}(q) \right\} \,\,, \nonumber \\
D_{12}(k)&=& - \,\, w^2_1(k) \sum_{{\bf q}} \frac{\left({\bf k} \cdot {\bf q} \right)^2}{k^3 q} \,\, h_{12}(|{\bf k} - {\bf q}|)  \,\,.
\end{eqnarray}

The poles of the inverse dielectric permeabilities determine the eigenfrequencies in the bilayer system. We find in accordance with \cite{KRG93,VKB96}
\begin{eqnarray} \label{QLCmodes}
\omega^2_{{\rm in}}(k)&=& w^2_{1}(k) (1+e^{-kd}) + (D_{11}+D_{12}) \,\,, \nonumber \\
\omega^2_{{\rm out}}(k)&=& w^2_{1}(k) (1-e^{-kd}) + (D_{11}-D_{12}) \,\,.
\end{eqnarray}
These expressions were analyzed in Refs. \cite{KRG93,VKB96}. It was found that the in-phase modes are not qualitatively different from the corresponding modes in the isolated 2D layer (with double density). In particular, for $k \to 0$ the typical 2D soft plasmon mode behavior $\omega \sim \sqrt{k}$ can be observed. On the contrary, from Eq.(\ref{QLCmodes}) one concludes that the out-of-phase modes develop an energy gap at $k=0$ \cite{KRG93,VKB96,KG98}:
\begin{equation} \label{gap}
\omega^2(0)= -\,\, \frac{e^2 n_S}{m} \int_0^{- \infty} \, dq \,\, q^2  e^{-qd}\, h_{12}(q) \,\,.
\end{equation}

However, this result looks quite strange. One expects that in the longwavelength limiting case $k \to 0$ the wave does not feel the separation between the two layers and we should deal with a double density single 2D layer. For the single layer, however, an out-of-phase mode is unknown. Moreover, as can be seen from Eq.(\ref{gap}) the energy gap for the out-of-phase mode remains even for the case of a vanishing layer separation $d=0$, i.e. for the case of a single 2D layer. Due to the translational invariance of the double density single 2D layer only the plasmon mode $\omega^2(k)=2w^2_1(k)$ should be observed as $k \to 0$. One concludes therefore that there should not be any energy gap in the bilayer system. Summarizing the above discussion, we have found that Eq.(\ref{QLCmodes}) predicts an energy gap which should not exist. 

To clarify this discrepancy we go back to Eqs.(\ref{QLCeps}). These equations describe the dielectric permeabilities of systems of oscillators with eigenfrequency  $\omega^2_{{\rm in}}(k)$ (or $\omega^2_{{\rm out}}(k)$) and with the oscillator strength $f_{{\rm in}}=1+e^{-kd}$ (or $f_{{\rm out}}=1-e^{-kd}$, respectively). We focus now on the out-of-phase mode and consider the long-wavelength limiting case $k \to 0$. The eigenfrequency of the out-of-phase mode reduces to the energy gap value (Eq.(\ref{gap})) but the oscillator strength tends to zero as $f_{{\rm out}}=kd$. This is just the solution of our problem. Within the QLC approach there is indeed a finite energy gap for the out-of-phase mode, but this mode does not develop as $k \to 0$ since the mode oscillator strength vanishes. In a quantummechanical language this means that the transition between two energy levels separated by an energy gap is forbidden; in a classical language one would say that the number of oscillators having the eigenfrequency $\omega^2(0)$ (Eq.(\ref{gap})) tends to zero as $k \to 0$. 

Our discussion results in the following picture within the QLC approach: as $k \to 0$ the eigenfrequency of the out-of-phase mode tends to the energy gap value but the mode signal becomes weaker and weaker. Now a deficancy of the QLC algorithm comes into play. The QLC method is not able to describe damping processes and therefore the calculations based on QLC approximation fail to provide information on the damping of the collective modes. However, with decreasing mode signal the noise to signal ratio increases and it is necessary to take into account damping processes.

In what follows we employ the classical method of moments to determine the dispersion law of the bilayer collective modes. To do so we use an interpolation formula, satisfying all known exact relations and sum rules. This allows us to include damping processes into our calculation, at least formally. In Ref. \cite{OT92} an expression for the dielectric permeability of a single 2D electron layer
satisfying all known sum rules is constructed. Generalizing the results of \cite{OT92} to the present case one expresses the dielectric permeability for the  out-of-phase motions via the frequency moments  $M_i=- \frac{1}{\pi}\int_{-\infty}^{\infty}\omega^i \, {\rm Im}\,\varepsilon^{-1}_{{\rm out}}({\bf {k}},\omega) \, d \omega$ ($i=-1,1,3$) and the function $q_{{\rm out}}=q_{{\rm out}}({\bf k},z)$ being analytic in the upper half-plane ${\rm Im}\,\,z >0\,,~z=\omega+i\,\,\eta$ and having a positive imaginary part there, and such that $[q_{{\rm out}}(k,z)/z] \to 0$ for $z \to \infty$,
\begin{eqnarray} \label{momentseps}
\varepsilon^{-1}_{{\rm out}}({\bf {k}},\omega)&=&1 + \frac{w^2_{1}(k) (1-e^{-kd})\,\, \left(\, \omega+q_{{\rm out}}({\bf k},\omega)\, \right)}{\omega \biggl( \omega^2 -w^2_{1}(k) (1-e^{-kd}) - (D_{11}-D_{12}) \biggr) + q_{{\rm out}} \biggl( \omega^2 - w^2_{1}(k) (1-e^{-kd})\left[1-\varepsilon^{-1}_{{\rm out}}(k)\right]^{-1}\biggr)} \,\,. 
\end{eqnarray}
Eq.(\ref{momentseps}) interpolates between the exact high-frequency behavior (given by the first and third frequency moment sum rules
and the low-frequency behavior (determined by $\varepsilon_{{\rm out}}(k)$ - the static dielectric permeability for the out-of-phase motion). For simplicity, in Eq.(\ref{momentseps}) we have omitted the kinetic energy term into the third frequency moment since we are interested in the strong coupling limit only. An expression similar to Eq.(\ref{momentseps}) can be obtained for the in-phase motion. Notice that the QLC expression Eq.(\ref{QLCeps}) can be obtained by putting $q_{{\rm out}} \equiv 0$. The QLC expression also satisfies the first and third frequency moment sum rules but does not reproduce the low-frequency behavior of the dielectric permeability. Via the inclusion of a finite (complex) function $q_{{\rm out}}$ one is able to describe the low-frequency behavior in the correct manner. In addition we have now the formal possibility to include damping into our considerations. In the liquid phase of a strongly coupled bilayer the damping is mainly given by the diffusion of the quasilocalized particles site positions. A nonzero diffusion constant for the site positions migration results in a nonzero conductivity for the out- and in-phase motions. We mention here that the 2D static ``conductivity'' $\sigma_{{\rm out}}$ is not a real conductivity since it describes the out-of-phase flow. Notice that the existence of a nonzero static out-of-phase ``conductivity `` is guaranteed only if $q_{{\rm out}}({ {k \to 0}}, 0) = i h$, with $h= 4 \pi \sigma_{{\rm out}} (D_{11}-D_{12}) / d_l w^2_{1}(k) (1-e^{-kd}) \, \sim k^{-2}$ ($d_l$ being an effective thickness of one layer).  We have no phenomenological way to choose the exact $q({\bf k},\omega)$. However, the most natural way to generalize the QLC result is to put $q_{{\rm out}}({\bf k}, \omega)$ equal to its static value in the long-wavelength limit $q_{{\rm out}}=ih$. Within this approximation the high frequency pole of the out-of-phase permeability is shifted into the lower complex half plane. Instead of Eq.(\ref{QLCmodes}) we have now:
\begin{eqnarray} \label{QLCmodesnow}
\omega^2_{{\rm out}}(k \to 0)&=&  -i \frac{h}{2} \pm \sqrt{- \frac{h^2}{4}+ (D_{11}-D_{12})} \,\,\,\,,
\end{eqnarray}
 and the mode corresponding to the energy gap in the QLC approximation becomes overdamped. 

In the previous discussion we have shown that the interlayer correlations lead to a divergence of the third to the squared first frequency moment ratio $M_3/M_1^2$ in the long-wavelength limit. Using the fact that $q_{{\rm out}} \sim M_3/M_1^2$ the expansion of Eq.(\ref{momentseps}) in terms of $M_1^2/M_3$ provides as $k \to 0$ the following expression:
\begin{eqnarray} \label{momentseps2}
\varepsilon^{-1}_{{\rm out}}({\bf {k} \to 0},\omega)&=&1 + \frac{w^2_{1}(k) (1-e^{-kd})}{ \omega^2 + \omega q'({\bf k},\omega) - w^2_{1}(k) (1-e^{-kd})\left[1-\varepsilon^{-1}_{{\rm out}}(k)\right]^{-1}} + O(k^2) \,\,. 
\end{eqnarray}
Here $q'({\bf k},\omega)=-(D_{11}-D_{12})/q_{{\rm out}}({\bf k},\omega)$ is a function which has the same properties as $q_{{\rm out}}({\bf k},\omega)$ has. Notice that Eq.(\ref{momentseps2}) contains the information on the static dielectric permeabilty $\varepsilon^{-1}_{{\rm out}}(k)$,which is given by a compressibility sum rule as $k \to 0$. The high-frequency part is given only by the first frequency moment, whereas the ``diverging'' third frequency moment is absent. Eq.(\ref{momentseps2}) corresponds therefore to a STLS-like approximation for the out-of-phase motion dielectric permeability. We have again no phenomenological way to choose the exact $q'({\bf k},\omega)$. However, from Eq.(\ref{momentseps2}) it can be seen that $\nu=-i\,q'({\bf k},\omega)$ can be regarded as an effective collision frequency in a Drude-Lorentz like theory. Neglecting the frequency dependence of this collision frequency for the out-of-phase motion one defines the collision frequency by the static ``conductivity'' of the out-of-phase motion \cite{OT92} $\sigma_{{\rm out}}$, $\nu(k)=w^2_{1}(k) (1-e^{-kd}) d_l /4\pi \sigma_{{\rm out}}$ . We will not perform a detailed analysis of this quantity. We mention here only that in a case of a finite out-of-phase ``conductivity'' we obtain that $\nu(k)$ behaves as $k^2$ for $k \to 0$. Then from Eq.(\ref{momentseps2}) we obtain the following dispersion for the out-of-phase motion in the long-wavelength limiting case,
\begin{equation} \label{final}
\omega^2_{{\rm out}}(k)=\frac{2 \pi e^2 n_s d}{m}\, k^2 + O(k^2) \,\,.
\end{equation}
This is just the result of the STLS calculations \cite{KG98}. Notice that Eq.(\ref{final}) corresponds to the low-frequency pole of Eq.(\ref{momentseps}). Thus we have found that the expression Eq.(\ref{momentseps}) calculated from the theory of moments contains the QLC expression and a STLS like expression as limiting cases. Further we have shown, that in the long wavelength limiting case Eq.(\ref{momentseps}) converts into a STLS like expression and does not predict the existence of an energy gap in the bilayer. Due to the absence of an energy gap in the present approach we observe a soft transition from the bilayer to the double density single layer with layer separation $d \to 0$. However, within the qualitative analysis of this paper we are not able to establish the value of wavevector $k$ at which the transition to the STLS like permeability occurs. However, some qualitative statements can be made. The ``transition point'' to the STLS regime is mainly given by the function $q_{{\rm out}} \sim \nu^{-1}(k)$. With increasing coupling strength the diffusion constant decreases and so does $\nu^{-1}(k)$. We also expect that the STLS ``transition point'' is shifted to lower $k$ values and it might be possible to observe an ``energy gap'' at not to low $k$ values.

In this Brief Report the dispersion law for the out-of-phase mode in an electronic bilayer is analyzed. The analysis is based on the expression for the dielectric permeability of the out-of-phase motion obtained from the classical theory of moments without using perturbation theory. We have shown that this expression contains the permeability expressions obtained by other approaches as the QLC or the STLS approximation as limiting cases. The analysis of our expression has confirmed the STLS predictions of an absence of an energy gap in the bilayer system. Finally, we invite the experimentalists to verify the different predictions of the various theoretical approaches concerning the existence of an energy gap in a bilayer. This would exhibit not only a check for this specific system but also a general verification of the various theoretical approaches applicable to other plasma systems. Another tool to investigate the collective mode structure in bilayers might be numerical simulations. However, due to the finite number of particles in the simulations the available $k$ values have a lower bound. Therefore one remains with the difficult task to extrapolate the behavior at $k=0$ from the behavior at relative high $k$ values. 
 
{\bf Acknowledgments.} Valuable discussions with G.~Kalman and I.~M.~Tkachenko are gratefully acknowledged. This work was financed by the Deutsche
Forschungsgemeinschaft.

\end{document}